\documentclass[aps,prl,twocolumn,showpacs,superscriptaddress]{revtex4-1}  
\usepackage{amsmath}
\usepackage{amssymb}
\usepackage{graphicx}
\usepackage{epstopdf}
\usepackage{suffix}
\usepackage{mathtools}
\usepackage{gensymb}

\begin{document}

\title{Electronic structure of carbon nanotubes on graphene substrates}

\author{Benedetta Flebus}
\address{Department of Physics and Astronomy, University of California, Los Angeles, California 90095, USA}
\author{Allan H. MacDonald}
\address{Department of Physics, The University of Texas at Austin, Austin, TX 78712, USA}

\begin{abstract}
Allotropes of carbon, including one-dimensional carbon nanotubes and two-dimensional graphene sheets, 
continue to draw attention as promising platforms for probing the physics of electrons in lower dimensions.  
Recent research has shown that the electronic properties of 
graphene multilayers are exquisitely sensitive to the relative orientation between sheets,
and in the bilayer case exhibit strong electronic correlations when close to a magic twist angle.  
Here, we investigate the electronic properties of a carbon nanotube
deposited on a graphene sheet by deriving a low-energy  theory that accounts both for rotations and rigid displacements of the nanotube with respect to the underlying graphene layer. We show that this heterostructure is described  by a translationally invariant, a periodic or a quasi-periodic Hamiltonian, depending on the  orientation and the chirality of the nanotube. Furthermore, we find that, even for a vanishing twist angle,
 rigid displacements of a nanotube with respect to a graphene substrate can  alter
its electronic structure qualitatively.  Our results identify a promising new direction for 
strong correlation physics in low dimensions.  
\end{abstract}



\maketitle

\textit{Introduction.} Carbon nanotubes and graphene sheets are, respectively, one and two dimensional carbon allotropes.
Both systems have been extensively studied for several decades because of their unique electrical, optical and mechanical 
properties~\cite{Ando,Guinea}, and their wide range of potential applications spanning  from
electric circuits to solar cells and exciton-polariton lasers~\cite{huang,avouris,euen,baumberg}. 
They have proven to be valuable platforms for investigating new physics in reduced dimensions, 
leading for instance to the first experimental observation of the integer quantum Hall effect at room 
temperature~\cite{Novoselov}.  Graphene bilayers 
share many properties with their monolayer counterparts~\cite{Novoselovbilayer}, 
but exhibit fascinating new phenomena as well~\cite{McCann}.  Notably, Cao \textit{et al.} and others have recently
observed a correlated insulator state~\cite{Cao1} and unconventional superconductivity~\cite{Cao2}  in graphene bilayers that 
have a relative orientation close to a magic~\cite{Rafi} angle near 1.1$\degree$, 
sparking excitement in the physics community.  Although they have not yet attracted wide attention, nanostructures formed between graphene and carbon nanotubes also hold promise since they might bring the sensitivity to orientations discovered in bilayers to one-dimensional physics.

\begin{figure}[b!]
\includegraphics[width=0.9\linewidth]{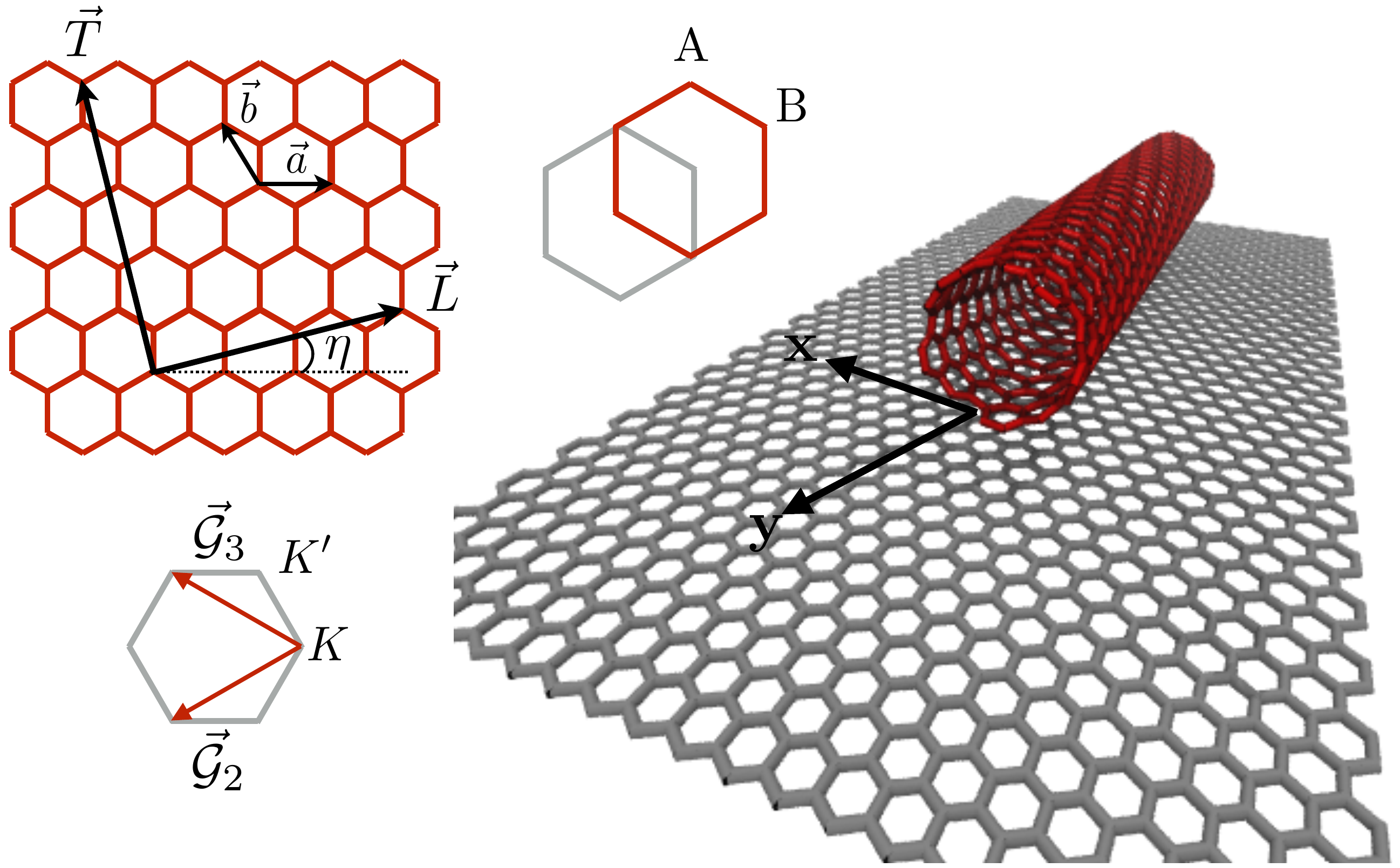}
\caption{\textbf{Single-wall carbon nanotube  on a graphene substrate.} 
The nanostructure geometry can be characterized by starting 
from Bernal AB stacking between a graphene sheet and an unrolled nanotube with general chirality, 
and then displacing the nanotube or changing the orientation of its axis.
In  Bernal stacking, atoms located at the B sites of the carbon nanotubes are above A sites of the graphene layer. 
Top left:  structure of an unrolled carbon nanotube, which is specified by a primitive translation 
vector $\vec{T}$ along the nanotube axis, and a perpendicular chiral vector $\vec{L}$.  Both vectors can be written as  linear combinations of the primitive lattice vectors $\vec{a}$ and $\vec{b}$.
Bottom left: $\vec{\mathcal{G}}_{2}$ and $\vec{\mathcal{G}}_{3}$ are reciprocal lattice vectors 
that connect equivalent graphene Brillouin-zone corners.  The momentum states 
near K and its equivalent corners together host the Dirac point of valley K.  
Momenta near the other three Brillouin-zone corners host valley $K'$, 
the time-reversed  counterpart of  valley $K$. 
Right: AB stacking between graphene and the unrolled nanotube applies 
when the orientation of the nanotube axis matches its chirality 
angle $\eta$.}
\label{Fig1}
\end{figure}

In this Letter, we establish a low-energy effective theory for hybridized nanotube$|$graphene heterostructures.  
We focus specifically on single-wall metallic nanotubes 
placed on a graphene layer, as illustrated in Fig.~\ref{Fig1}. We derive a Hamiltonian describing electron tunneling between a graphene layer and a nanotube deposited at a generic twist angle.
We find that, even for vanishing twist angle, rigid displacements of the nanotube  with respect to the graphene sheet strongly alter the nanotube electronic properties, 
leading, e.g., to valley-dependent differences between the
Dirac velocities of left and right movers, similar to the spin-dependent velocities 
induced in quantum wires by Rashba spin-orbit interactions~\cite{carr}.  
For some magic displacements, nanotube group velocities in the vicinity of the Dirac points
are strongly reduced, suggesting that strong interaction regimes and possibly~\cite{volovik}
superconducting transitions might be made accessible by simple mechanical displacements.
%

\textit{Model.} We consider a single-wall metallic nanotube deposited on top of a graphene layer, as 
illustrated in Fig.~\ref{Fig1}.  The nanotube geometry  is uniquely determined by the
translational vector $\vec{T}=m_{a} \vec{a} + m_{b} \vec{b}$, oriented along the nanotube axis,
and the chiral vector $\vec{L}=n_{a} \vec{a} + n_{b} \vec{b}$, with $L=|\vec{L}|$ being the nanotube circumference.
Here, $\vec{a}$ and $\vec{b}$ are the primitive vectors of the graphene sheet's triangular lattice illustrated in Fig.~\ref{Fig1}.
The sets of integers $[n_{a,b}, m_{a,b}]$ are constrained by the orthogonality
condition $\vec{L} \cdot \vec{T}=0$.  Metallic nanotubes occur when the condition $n_{a} + n_{b}=3 N +1$
is satisfied, with $N$ being an integer.
For nanotubes with diameter larger than a few nanometers, we can safely neglect the effects of curvature of its wrapped
graphene sheet~\cite{Ando, Millie}.  In this case,
the nanotube electronic states can be obtained by imposing the boundary condition 
$\psi(\vec{r})=\psi(\vec{r}+\vec{L})$ on the graphene wave functions,
resulting in the momentum quantization condition 
$p_{x}\equiv p_{x}(j)=2 \pi j/ L$~\cite{Ando}, with $j$ integer. 
We focus on the limit in which the interlayer tunneling strength is weaker 
than the energy separation between nanotube minibands, allowing us to 
truncate the low-energy Hilbert space to the $j=0$ subspace. 

We consider an AB-stacked arrangement, as shown in Fig.~\ref{Fig1}, which can be achieved when the nanotube orientation on the 
graphene sheet matches the tube chirality angle $\eta$.  The atomic positions in the graphene layer and the nanotube can be written, respectively, 
as $\vec{R}$ and $\vec{R}'=M(\theta) (\vec{R}-\tau)+\vec{d}$, where the matrix $M(\theta)$ describes left-handed rotation of angle $\theta$, $\tau$ is a vector connecting the two atoms in the unit cell and $\vec{d}$ is a translation vector. In our coordinate system,  the nanotube axis is oriented along the $y$ direction. With these assumptions,  the Dirac Hamiltonians of the carbon nanotube, $h(p_{y})$,
and the graphene layer, $h_{g}(\vec{k},\eta,\theta)$, can be written as
\begin{align}
h(p_{y})=&-v\begin{bmatrix} 0 &  i p_{y} \\ -ip_{y} & 0\end{bmatrix}\,, \nonumber \\
h_{g}(\vec{k}, \eta, \theta)=&-v\begin{bmatrix} 0 & k^{+} e^{-i (\eta - \theta)} \\ k^{-} e^{i (\eta- \theta) } & 0\end{bmatrix}\,,
\end{align}
where $\vec{k}$ (with $k^{\pm}=k_{x} \pm i k_{y}$) and $p_{y}$ are momenta measured with respect to the respective Dirac points 
and $v$ is the Dirac velocity.  

Here, we assume that the overlap between the $\pi$ orbitals
of the two subsystems can be represented by a function $t(\vec{r})$ that is smooth on the scale of the lattice spacing~\cite{Rafi}.  For the explicit calculation, we adopt a two-center approximation for the interlayer hopping
amplitude~\cite{Rafi} and account for the finite transverse size of the nanotube by assuming that  
the hopping amplitude  varies with position in the $\vec{L}$ direction~\cite{footnote}.  For definiteness, we take 
$t(x) \propto  e^{-(x/L)^2}/L$.  

The tunneling matrix element 
describing a process in which an electron in valley $K$ with momentum $p_{y}$ residing on sublattice $\beta$ 
of the nanotube hops to momentum state $\vec{k}$ and sublattice $\alpha$ in the same valley of the graphene layer is then~\cite{supplementary}
\begin{align}
T_{\vec{k}, p_{y}}^{\alpha \beta}=\frac{t}{\sqrt{2\pi}} \sum_{j=1}^{3} T^{\alpha \beta}_{j} \; \text{exp}\left[ -\frac{L^2}{2}(k_{x}-q_{jx})^2 \right] \delta_{k_{y}-p_{y},q_{jy}}\,,
\label{tunneling}
\end{align}
where  $t\equiv t(k_{D})$ is the Fourier transform of the tunneling amplitude $t(\vec{r})$ and $k_{D}$ is the magnitude of the Brillouin-zone corner wave vector. Here, we have introduced the vectors $\vec{q}_{j}=k_{\theta} \left( \sin(\chi_{j}+\theta), - \cos(\chi_{j} +\theta)  \right)$, with $k_{\theta}=2k_{D}\sin(\theta/2)$, and $\chi_{j}=2\pi (j-1) /3$, and  the tunneling matrices $T_{j}$, i.e.,
\begin{align}
& T_{1}= \begin{pmatrix} 1 & 1 \\ 1 & 1 \end{pmatrix}\,, \; \; \;  T_{2}=e^{- i \vec{\mathcal{G}}_{2} \cdot \vec{d}}\begin{pmatrix} e^{-i \phi} & 1 \\ e^{i \phi} & e^{-i \phi} \end{pmatrix}\,, \\ \nonumber 
&\text{and} \; \;  T_{3}=e^{- i \vec{\mathcal{G}}_{3} \cdot \vec{d}}\begin{pmatrix} e^{i \phi} & 1 \\ e^{-i \phi} & e^{i \phi} \end{pmatrix}\,.
\end{align}
Here,  $\phi=2\pi/3$ and $\vec{\mathcal{G}}_{2,3}$  are the reciprocal lattice vectors depicted in Fig.~\ref{Fig1}. For a vanishing twist angle, the tunneling Hamiltonian~(\ref{tunneling}) is translationally invariant along the $y$ direction and the $y$-component of the momentum is a good quantum number. For a twist angle $\theta=\arctan(q/\sqrt{3})$, with $q$  a rational number, the $y$-component of the momentum is a good quantum number up to a reciprocal lattice vector~\cite{supplementary}. The corresponding Hamiltonian is then periodic along the $y$ direction and can be diagonalized in terms of one-dimensional Moir\'e bands using a plane wave expansion.  
For other twist angles,  the spectrum is quasi-periodic: the corresponding  wave functions are quasi-localized and  tunneling is suppressed~\cite{quasiperiodic}.

\begin{figure*}[htbp]
    \centering
    \includegraphics[width=0.92\linewidth]{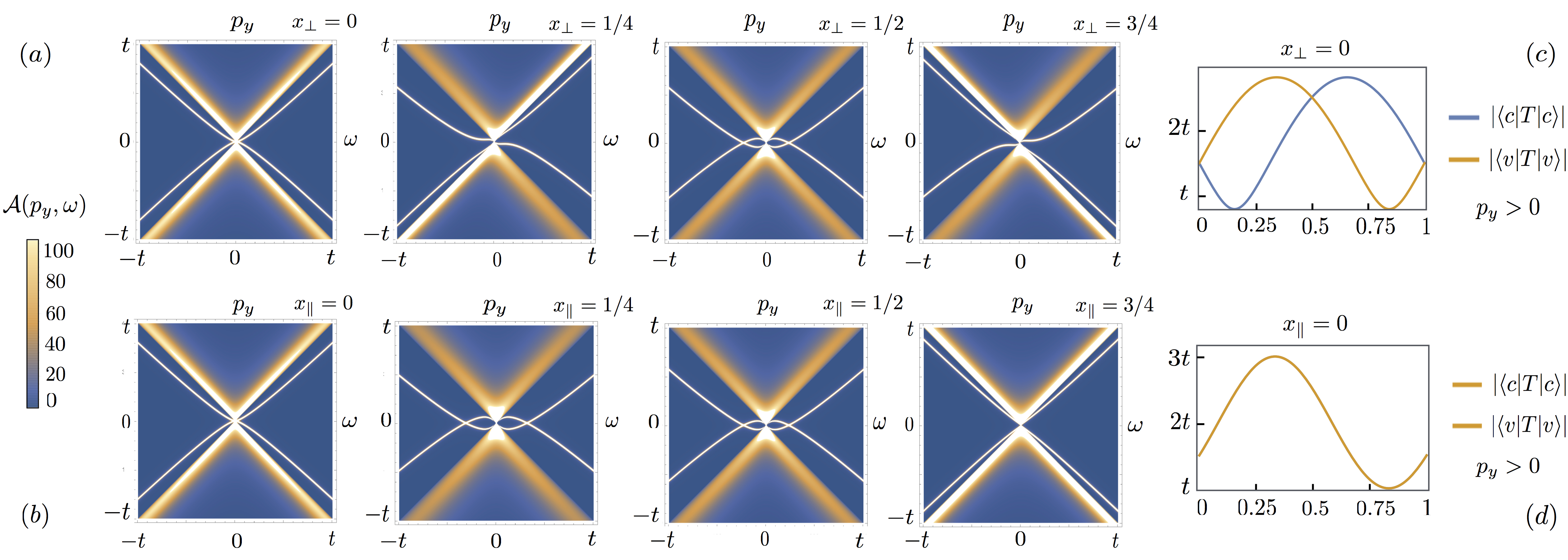}
    \caption{\textbf{Electronic structure of a carbon nanotube on a graphene sheet.} Spectral function $\mathcal{A}(p_{y},\omega)$ of the carbon nanotube for rigid displacements along the direction transverse (a) and parallel (b) to the nanotube axis. (c) Dependence of the (absolute value of) conduction
 ($|\langle c|T|c\rangle|$) and valence ($|\langle v|T|v\rangle|$) band hybridization matrix elements~(\ref{94}) at the graphene momentum $k_x=0$ on displacement
 transverse to the nanotube axis. For the dimensionless displacements $x_{\bot} \neq 0, 1/2$, the two matrix elements have different values, explaining 
 the electron-hole asymmetry of the spectral function (a). (d) For displacements along the nanotube axis, the 
 valence and conduction bands experience the same interaction strength, explaining
 the electron-hole symmetry of the spectral function (b). The spectral densities, frequencies, momenta and tunneling strength are plotted in dimensionless 
units, {\it i.e.} $\mathcal{A} \rightarrow (v/a) \mathcal{A} $, $ p_{y} \rightarrow a p_{y} $, $\omega \rightarrow \omega (a/v)$ and $t \rightarrow t  (a/v)$.}
    \label{Fig2}
\end{figure*}

In this Letter, we focus on the translationally-invariant case, i.e.,  $\theta=0$, showing that even a rigid displacement can strongly affect nanotube electronic properties. 
Below we characterize how the  hybridization between the nanotube and the graphene sheet alters the nanotube
electronic properties by evaluating full momentum- and frequency-resolved spectral function~\cite{bruus} of the nanotube:
\begin{align}
\mathcal{A}(p_{y},\omega)=-\frac{1}{\pi}  \sum_{\gamma} \text{Im} G^{R}_{ \gamma \gamma}(p_{y},\omega)\,,
\label{spectral}
\end{align}
where $G^{R}_{ \gamma \gamma}$ is the $\gamma \gamma$th matrix element of the nanotube retarded Green's function, i.e., 
\begin{align}
 G^{R}(p_{y}, \omega)=\lim_{\epsilon \rightarrow 0}\left[\left( \omega + i \epsilon \right) \mathcal{I} - h(p_{y}) - \Sigma (p_{y}, \omega) \right]^{-1}\,.
\end{align}
Here, $\mathcal{I}$ is the $2 \times 2$ identity matrix, while
\begin{align}
\Sigma_{ij} (p_{y}, \omega)=\int dk_{x} \left[ T^{\dagger}(p_{y}) G^{R}_{g}(k_{x}, p_{y}, \eta, \omega) T(p_{y})  \right]_{ij}\,
\end{align}
is the $ij$th  matrix element of the carbon nanotube self-energy due to hybridization with the graphene sheet.  Here, $G_{g}^{R}$  is the retarded Green's function of the unperturbed graphene sheet~\cite{supplementary}.  

\textit{Results.} In this section, we present the results of a numerical evaluation of  the spectral 
density~(\ref{spectral}) as function of a rigid displacement $\vec{d}$ of the nanotube.
We distinguish translations transverse, $\vec{d}_{\bot}$, and 
parallel, $\vec{d}_{\parallel}$, to the nanotube axis and define dimensionless translations $x_{\bot (\parallel)}$ by $\vec{d}_{\bot}=\sqrt{3} a x_{\bot} (0, 1)$ and $\vec{d}_{\parallel}= a x_{\parallel} (1, 0)$, with $a$ being the lattice constant of graphene. The electronic structure is a periodic function of $x_{\bot (\parallel)}$ with period $1$.  
Results for several different
values of $x_{\parallel (\bot)}$ at $t=100$ meV, $L=10 \pi$ nm, and  $\eta=0$, 
are summarized in Figs.~\ref{Fig2}(a) and (b).
Figure~\ref{Fig2}(a) shows the dependence of the spectral density~(\ref{spectral}) on translations transverse to the nanotube axis. 
The isolated bands illustrate the strong influence of hybridization on the nanotube band dispersion, while
the continuous spectra reflect leakage of the graphene sheet orbitals onto the nanotube.  
We observe that the nanotube Dirac velocities are generically reduced by level repulsion with graphene sheet 
orbitals and that the  linear dispersion of isolated nanotubes can be strongly distorted.  
For a general translation $x_{\bot} \in (0,1)  \neq 0, 1/2$, 
the nanotube spectrum displays an asymmetry between right-goers and left-movers,
and the graphene spectral weight shows a corresponding asymmetry.
 On the other hand, as illustrated in Fig.~\ref{Fig2}(b),
symmetric spectra and spectral weight are maintained
for any translation along the nanotube axis.  

Most features of Figs.~\ref{Fig2}(a) and (b) can be understood  by 
examining the hybridization matrix elements between the nanotube and graphene conduction and valence bands, while  setting the $x$-component of the graphene momentum $k_x \to 0$. 
This suffices to capture the largest contribution to the spectral density, which is due to energy-conserving processes 
in which electrons tunnel from the nanotube to the graphene layer and {\it vice versa}. 
Since the hopping term~(\ref{tunneling}) preserves the $y$-component of  momentum, 
energy is conserved during tunneling only when $k_{x}=0$. 
The hybridization matrix elements 
between graphene and nanotube conduction and 
valence bands at $k_x=0$ are 
\begin{align}
T_{cc}\equiv \langle c|T|c \rangle&=\frac{1}{2} \left[ T^{AA} +T^{BB} + i \left( T^{AB} - T^{BA}\right) \right]\,,  \nonumber \\
T_{vv}\equiv \langle v|T|v \rangle&=\frac{1}{2} \left[ T^{AA} +T^{BB} + i \left( T^{BA} - T^{AB}\right) \right]\,, 
\label{94}
\end{align}
for $p_{y} >0$, while for $p_{y}<0$ $T_{cc} \leftrightarrow T_{vv}$.
Figures~\ref{Fig2}(c) and (d)  illustrate the variation of the matrix elements in Eq.~(\ref{94}) 
upon transverse and parallel displacement, respectively.
Considering, for instance, $x_{\bot}=3/4$ and $p_{y}>0$, one finds that the repulsive interaction 
between conduction bands is much larger than the one between valence bands, i.e., $|T_{cc}| \gg | T_{vv}|$. 
As a result, the velocity renormalization of the nanotube conduction band 
is larger than for the valence band.  Larger velocity reductions also lead to smaller amplitudes 
for the graphene sheet orbitals on the nanotube.  
\begin{figure*}[htbp]
    \centering
    \includegraphics[width=0.92\linewidth]{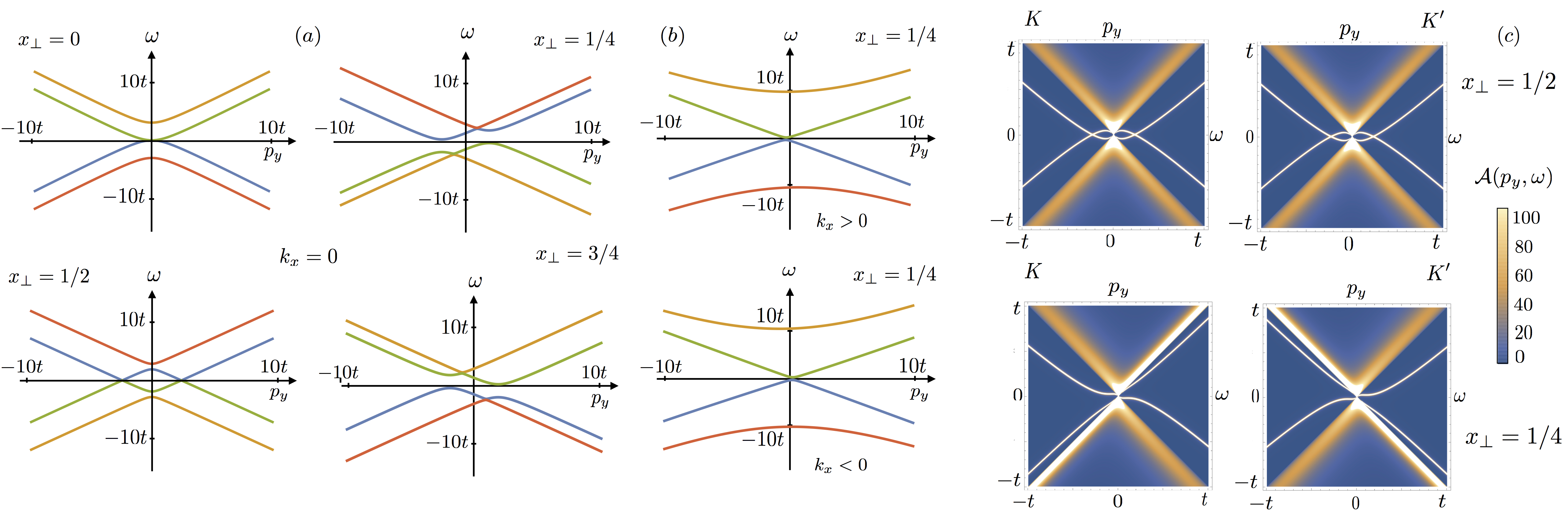}
    \caption{\textbf{Four-band model and valley-dependent spectral functions.} (a) Energy spectrum of the $4 \times 4$ Hamiltonian~\eqref{101} for graphene momentum $k_{x}=0$ and different values of dimensionless displacement $x_{\bot}$.  Far  from the Dirac point, the spectrum displays the same features observed in the spectral function, see Fig.~\ref{Fig2}(a). However, in contrast with the spectral function, the four-band spectrum exhibits an energy gap.  (b) Energy spectrum of the $4 \times 4$ Hamiltonian~\eqref{101} for a finite  $k_{x}$ and $x_{\bot}=1/4$. For $k_{x} >0$ $(<0)$, the gap closes at the left (right) of the Dirac point. The total contribution of electronic states with finite $k_{x}$ gives rise to a gap closing at the Dirac point, which is displayed by the spectral density in Fig.~\ref{Fig2}(a).  (c) Nanotube spectral density $\mathcal{A}(p_{y},\omega)$ at the valleys $K$ and $K'$, which are related by time-reversal symmetry. An electron-hole symmetric spectrum is  degenerate for $K$ and $K'$ (top figure). When electron-hole symmetry is broken (bottom figure), a right-going state at momentum $p_y$ in one valley still has a left-going degenerate Kramers partner at momentum $-p_y$ in the other valley.   The spectral densities, frequencies, momenta and tunneling strength are plotted in dimensionless 
units, {\it i.e.} $\mathcal{A} \rightarrow (v/a) \mathcal{A} $, $ p_{y} \rightarrow a p_{y} $, $\omega \rightarrow \omega (a/v)$ and $t \rightarrow t  (a/v)$. }
    \label{Fig3}
\end{figure*}

Although the strongest trends in 
Fig.~\ref{Fig2}(a) can be understood by examining level repulsion separately for 
conduction and valence bands, the behavior near $p_y=0$ involves all four bands.
The full four-band Hamiltonian of the coupled bilayers is 
\begin{align} 
\mathcal{H}=\begin{pmatrix} 0 && h_{g 12} && T_{11} && T_{12} \\ h_{g 21} && 0 && T_{21}   && T_{22} \\ T^{*}_{11} && T^{*}_{21} && 0 && h_{12} \\ T^{*}_{12} && T^{*}_{22} && h_{ 21} && 0 \end{pmatrix}\,.
\label{101}
\end{align}
Figure~3(a) shows that diagonalizing Eq.~(\ref{101}) at $k_{x}=0$ explains most of the behavior seen 
Fig.~\ref{Fig2}(a), although some gaps in the four band spectrum do not survive in the nanotube spectral function.
The absence of such gap can be explained by accounting for graphene states with $k_x \ne 0$. Figure~\ref{Fig3}(b) shows that diagonalizing Eq.~(\ref{101}) for a finite $k_x$ leads to a gap closing, respectively at the left and 
right of the Dirac point for $k_{x}>0$ and $k_{x}<0$.   Adding these two contributions 
results into a gap closing at the Dirac point, i.e.,  $p_{y}=0$. Finite momentum contributions lead also to a shift of the graphene-like 
conduction and valence bands to higher energies (in absolute magnitude). This is reflected by the spectral weight 
broadening we observe in correspondence of the graphene Dirac cone.


Another interesting feature of our results is the broken valley degeneracy of the spectrum, which we observe for any transverse translation (with the exception of $x_{\bot}=0,1/2$), but never when considering  displacements parallel to the nanotube axis. This property of the spectrum follows from the symmetries of our model. Time-reversal symmetry constrains the spectrum 
at momentum $p_{y}$ in the K-valley to equal the spectrum in the $K'$ valley at momentum $-p_{y}$,
as seen in Fig.~\ref{Fig3}(c).  When the valley-projected spectrum  is electron-hole symmetric, the two valleys are degenerate at each $p_y$. 
When  electron-hole symmetry is absent, on the other hand,
the spectrum of the $K$ valley can be mapped into the $K'$ valley by setting $p_{y} \rightarrow - p_{y}$.

 \textit{Discussion and outlook.}
In this work, we have established a low-energy effective model for a carbon nanotube on top of a graphene layer, which is valid for any displacement or rotation of the nanotube axis. Depending on tube orientation and chirality, the heterojunction Hamiltonian can be translationally invariant, periodic, or quasiperiodic.  In the translationally invariant case, we show that, even at a vanishing twist angle, rigid displacements of the nanotube with respect to the graphene layer can strongly alter the electronic properties of the former. 
 For instance,  a rigid displacement of the nanotube can break the particle-hole symmetry in the valley-projected spectrum, lift the degeneracy between the two valleys, and strongly alter the carrier velocity. 
 
These features, which can be explained through a four-band model and symmetry considerations, might be probed experimentally by NanoARPES, Raman spectroscopy or momentum and energy resolved tunneling spectroscopy \cite{ashoori}.   
Controlled growth or deposition of nanotubes on a substrate has been already demonstrated~\cite{xiao,gramich} and nanotubes can be laterally displaced on top of a graphene layer by, e.g., atomic force microscopy  \cite{seydou}, allowing our predictions to be tested experimentally.  

Our work suggest that atomic force manipulation of carbon nanotubes on graphene substrates can radically alter electronic properties, leading in some cases to strong correlations related to flattened nanotube bands, and to interesting modified electronic structures with radically different quasiparticle velocities in different valleys.  

Finally, we have assumed that the nanotube does not experience radial or axial deformation when lying on a substrate~\cite{yakob}. Such deformations can give rise, for instance,  to a lattice mismatch with respect to the graphene substrate or to the emergence of a gap in the nanotube dispersion~\cite{yang}. As such deformations have been experimentally observed~\cite{hertel,dai,minot}, future work should address them systematically.
\\
\\
\textit{Acknowledgements.} The authors thank Y. Yuval for insightful discussions. B.F. was supported by the Dutch Science Foundation (NWO) through a Rubicon grant. A. H. MacDonald was supported by 
DOE grant DE-FG02-02ER45958 and 
Welch Foundation Grant TBF1473.


\begin{thebibliography}{99}
\bibitem{footnote} In practice, the fall off of the tunneling strength as a function of distance from the center of the nanotube/graphene contact region will depend on nanotube shape deformation, but these details do not alter our results in an important way.
\end{thebibliography}
\end{document}


\title{Supplemental Material: Electronic structure of carbon nanotubes on graphene substrates}

\author{Benedetta Flebus}
\address{Department of Physics and Astronomy, University of California, Los Angeles, California 90095, USA}
\author{Allan MacDonald}
\address{Department of Physics, The University of Texas at Austin, Austin, TX 78712, USA}

\maketitle
In this supplemental material, we provide the derivation of the matrix element describing electron tunneling between a carbon nanotube and a graphene layer underneath it, i.e., Eq.~(3) of the main text. Moreover, we discuss the commensurability condition for the nanotube$|$graphene heterostructure and we derive the explicit form of the nanotube self-energy matrix. \\
\textit{Tunneling matrix and commensurability.} Our starting point is the tunneling matrix element between two graphene layers, which can be written as
\begin{align}
\tilde{T}^{\alpha \beta}_{\mathbf{k} \mathbf{p}}= t \sum_{j=1}^3 T^{\alpha \beta}_{j} \delta_{\mathbf{k}- \mathbf{p}, \mathbf{q}_{j}}\,.
\label{62}
\end{align}
The Fourier transform of Eq.~(\ref{62})  with respect to the variable $\mathbf{j}=\mathbf{k}-\mathbf{p}$ reads as
\begin{align}
\tilde{T}^{\alpha \beta}(\mathbf{r})=t \int \frac{d\mathbf{j}}{(2\pi)^2} \; e^{i \mathbf{j} \cdot \mathbf{r}}  \sum_{j=1}^{3}  \; T^{\alpha \beta}_{j} \delta_{\mathbf{j},  \mathbf{q}_{j}}\,.
\label{67}
\end{align}
In order to account  for the nanotube finite transverse size, we introduce the following spatial cut-off:
\begin{align}
T^{\alpha \beta}(\mathbf{r})= \frac{1}{L} e^{-\frac{x^2}{2L^2}} \tilde{T}^{\alpha \beta}(\mathbf{r})\,.
\label{71}
\end{align}
By Fourier transforming Eq.~(\ref{71}) while using Eq.~(\ref{67}), we obtain
\begin{align}
T^{\alpha \beta}_{\mathbf{k} \mathbf{p}}=\frac{t}{\sqrt{2\pi}} \sum_{j=1}^{3} T^{\alpha \beta}_{j} \; e^{ -\frac{L^2}{2}(k_{x}-p_{x}-q_{jx})^2 } \delta_{k_{y}-p_{y},q_{jy}}\,,
\label{77}
\end{align}
where $\vec{q}_{j}=k_{\theta} \left( \sin(\chi_{j}+\theta), - \cos(\chi_{j} +\theta)  \right)$, with $k_{\theta}=2k_{D}\sin(\theta/2)$ and $\chi_{j}=2\pi (j-1) /3$. 
Focusing on  the nanotube first miniband, i.e., $p_{x}=0$, allows us to further simplify Eq.~(\ref{77}) as
\begin{align}
T^{\alpha \beta}_{\mathbf{k} p_{y}}=\frac{t}{\sqrt{2\pi}} \sum_{j=1}^3 T^{\alpha \beta}_{j} \; e^{ -\frac{L^2}{2}(k_{x}-q_{jx})^2 } \delta_{k_{y}-p_{y},q_{jy}}\,.
\label{78}
\end{align}
For a vanishing twist angle, i.e., $\theta=0$, Eq.~(\ref{78}) reduces to
\begin{align}
T^{\alpha \beta}_{\mathbf{k} p_{y}}=\frac{t}{\sqrt{2\pi}} \sum_{j=1}^3 T^{\alpha \beta}_{j} \; e^{ -\frac{( k_{x} L)^2}{2} } \delta_{k_{y},p_{y}}\,.
\label{79}
\end{align}
Equation~(\ref{79}) shows that, while the translational invariance is broken along the $x$ direction,  the  $y$-component of the nanotube momentum is preserved upon tunneling, i.e., the latter is a good quantum number.

For a finite twist angle, the system is commensurate when the ratio between the vectors $\vec{q}_{j}$ is rational, i.e., 
\begin{align}
\frac{\cos \theta}{\cos (\theta + \chi_{1})}=\frac{n_{1}}{m_{1}}\,, \; \; \; \; \; \; \;  \frac{\cos \theta}{\cos (\theta + \chi_{2})}=\frac{n_{2}}{m_{2}}\,,
\label{95}
\end{align}
where $n_{1,2}$ and $m_{1,2}$ are integers. Equation~(\ref{95}) can be rewritten as
\begin{align}
\frac{n_{1,2}}{m_{1,2}}=-\frac{1}{2} \mp \frac{\sqrt{3}}{2} \tan \theta\,.
\label{100}
\end{align}
The condition~(\ref{100}) is satisfied when $\theta=\arctan ( q/\sqrt{3})$, with $q$ being a rational number. In this case, the $y$-component of the momentum is a good quantum number module a reciprocal lattice vector with $y$-component $G_{y}=k_{\theta} \cos \theta/(2\pi n)$, with $n=\text{min}(n_{1},n_{2})$. When the condition~(\ref{100}) is not fullfilled, the Hamiltonian becomes quasi-periodic.

\textit{Spectral function calculation.}
By solving
\begin{align}
\left[ i \mathcal{I} \frac{\partial}{\partial t} -h_{g}(\vec{k},\eta, \theta)  \right] G^{R}_{g}(\vec{k},t)&= \mathcal{I} \delta(t-t')\,,
\end{align}
while setting $\eta, \theta=0$, 
 the matrix elements of the retarded Green's function of the unperturbed graphene layer can be found  as
\begin{align}
&G^{R}_{g,11(22)}(\vec{k}, \omega)= \frac{\omega}{(\omega+i\epsilon)^2 - (v k)^2 }\,, \nonumber \\
&G^{R}_{g,21(12)}(\vec{k}, \omega)= \frac{ v k_{x}  \mp i v k_{y}}{ (vk)^2-(\omega+i\epsilon)^2 }  \,,
\end{align}
with $\epsilon \rightarrow 0^{+}$. We introduce an auxiliary matrix $\tilde{\boldsymbol{\Sigma}} \propto \boldsymbol{\Sigma}$, whose  $ij$ element is defined as 
\begin{align}
\tilde{\Sigma}_{ij}(p_{y},\omega)= \delta_{k_{y},p_{y}} \int dk_{x} \; e^{-\frac{k^2_{x} R^2}{2}} G^{R}_{g,ij}(k_{x},k_{y},\omega)\,.
\label{auxiliaryself}
\end{align}
Equation~(\ref{auxiliaryself}) can be recasted as $\tilde{\boldsymbol{\Sigma}}=\tilde{\boldsymbol{\Sigma}}^{r}+ i \tilde{\boldsymbol{\Sigma}}^{d}$, where $\tilde{\boldsymbol{\Sigma}}^{r(d)}$ represents the reactive (dissipative) contribution to the matrix $\tilde{\boldsymbol{\Sigma}}$. 
For $ |\omega|>v |p_{y}|$, the components of the dissipative part of Eq.~(\ref{auxiliaryself}) can be written as
\begin{align}
\tilde{\Sigma}^{d}_{11(22)}(p_{y},\omega)=&-  \frac{\pi |\omega| }{v} \frac{e^{[p^2_{y}-(\omega/v)^2] R^2}}{\sqrt{\omega^2-p_{y}^2 v^2}}\,,\label{142} \\
\tilde{\Sigma}^{d}_{21(12)}(p_{y},\omega)=&\mp  i \pi    p_{y}  \frac{ \omega}{|\omega|}  \frac{e^{[p^2_{y}-(\omega/v)^2] R^2}}{\sqrt{\omega^2 - p^2_{y} v^2}}\,, \label{143}
\end{align}
while, for $ |\omega|<v |p_{y}|$, they both vanish. For  $|\omega|<v |p_{y}|$, the matrix elements of the reactive contribution read as 
\begin{align}
&\tilde{\Sigma}^{r}_{11(22)}(p_{y},\omega)=- \pi^2 \frac{ \omega}{v} \frac{e^{[p_{y}^2- (\omega/v)^2 ] R^2}}{ \sqrt{ (v p_{y})^2-\omega^2}}  E \,, \nonumber \\
&\tilde{\Sigma}^{r}_{21(21)}(p_{y},\omega)=\pm i \pi^2 p_{y}   \frac{e^{[p^2_{y}-(\omega/v)^2 ]R^2} }{ \sqrt{ (v p_{y})^2-\omega^2}}  E\,, \nonumber \\
&\text{with} \; \; E\equiv E \left(\sqrt{p_{y}^2-\frac{\omega^2}{v^2}}  R\right)\,.
\label{149}
\end{align}
Here, we have introduced the function $E(x)=\frac{2}{\sqrt{\pi}} \int^{x}_{0} e^{-y^2} \; dy$. For $ |\omega|>v |p_{y}|$, we have instead
\begin{align}
&\tilde{\Sigma}^{r}_{11(22)}(p_{y},\omega)=\frac{2\sqrt{\pi}(\omega/v)}{\sqrt{\omega^2-(v p_{y})^2}}  F \,, \nonumber \\
&\tilde{\Sigma}^{r}_{21(12)}(p_{y},\omega)=\pm \frac{2 i\sqrt{\pi} p_{y}}{\sqrt{\omega^2-(v p_{y})^2}}  F \,, \nonumber \\
&\text{with} \; \; F\equiv F\left( \sqrt{ \frac{\omega^2}{v^2}-p^2_{y}} R \right)\,,
\label{155}
\end{align}
with $F(x)=e^{-x^2} \int^{x}_{0} e^{y^2} \; dy$. 
Equations~(\ref{149}) and (\ref{155}) have been tested numerically by invoking the Kramers-Kronig relations for the dissipative components~(\ref{142}) and (\ref{143}). \\